# Oxygen and Cation Ordered Perovskite, $Ba_2Y_2Mn_4O_{11}$

M. Karppinen[1,]*, H. Okamoto[1], H. Fjellvåg[1,2], T. Motohashi[1] and H. Yamauchi[1]

[1]*Materials and Structures Laboratory, Tokyo Institute of Technology, Yokohama 226-8503, Japan*
[2]*Department of Chemistry, University of Oslo, Blindern N-0315 Oslo, Norway*

A three-step route has been developed for the synthesis of a new oxygen-ordered double perovskite, $BaYMn_2O_{5.5}$ or $Ba_2Y_2Mn_4O_{11}$. (*i*) The *A*-site cation ordered perovskite, $BaYMn_2O_{5+\delta}$, is first synthesized at $\delta \approx 0$ by an oxygen-getter-controlled low-$O_2$-pressure encapsulation technique utilizing FeO as the getter for excess oxygen. (*ii*) The as-synthesized, oxygen-deficient $BaYMn_2O_{5.0}$ phase is then readily oxygenated to the $\delta \approx 1$ level by means of 1-atm-$O_2$ annealing at low temperatures. (*iii*) By annealing this fully-oxygenated $BaYMn_2O_{6.0}$ in flowing $N_2$ gas at moderate temperatures the new intermediate-oxygen-content oxide, $BaYMn_2O_{5.5}$ or $Ba_2Y_2Mn_4O_{11}$, is finally obtained. From thermogravimetric observation it is seen that the final oxygen depletion from $\delta \approx 1.0$ to 0.5 occurs in a single sharp step about 600 $^oC$, implying that the oxygen stoichiometry of $BaYMn_2O_{5+\delta}$ is not continuously tunable within $0.5 < \delta < 1.0$. For $BaYMn_2O_{5.5}$ synchrotron x-ray diffraction analysis reveals an orthorhombic crystal lattice and a long-range ordering of the excess oxygen atoms in the $YO_{0.5}$ layer. The magnetic behavior of $BaYMn_2O_{5.5}$ (with a ferromagnetic transition at ~133 K) is found different from those previously reported for the known phases, $BaYMn_2O_{5.0}$ and $BaYMn_2O_{6.0}$.

*Corresponding author:
Prof. Maarit Karppinen
Materials and Structures Laboratory
Tokyo Institute of Technology
4259 Nagatsuta, Midori-ku
Yokohama 226-8503, JAPAN
fax: +81-45-924-5365
e-mail: karppinen@msl.titech.ac.jp

## 1. Introduction

Cation-ordered perovskite oxides are promising candidates for novel functional materials. The structure of such oxides is derived from that of the simple perovskite, $ABO_3$, upon co-occupation of either the *A* or the *B* cation site with more than one metal species of different charges and/or sizes. Ordering at the *A* site is typically achieved with divalent Ba (*A* cation) and trivalent rare earth element, *RE* (*A*' cation). Moreover, oxygen-deficiency around the smaller *A*' cation is common. By varying the *A*:*A*' ratio (while keeping the (*A*+*A*'):*B* ratio at 1:1), it is possible to build up layered perovskite structures consisting of *A*O, $A'O_\delta$ and $BO_2$ layers. For the *A*:*A*' ratio of 1:1 and 2:1, oxygen-deficient $AA'B_2O_{5+\delta}$ double perovskite (DP) and $A_2A'B_3O_{6+\delta}$ triple perovskite (TP) structures are formed. Of the exciting phenomena so far revealed for such *A*-site ordered perovskites one may select as examples high-$T_c$ superconductivity (*e.g.* $Ba_2RECu_3O_{6+\delta}$ (TP) with $0.5 < \delta < 1.0$ [1]), magnetoresistivity (*e.g.* Ba(Eu/Gd)$Co_2O_{5+\delta}$ (DP) [2]), metal-insulator transition (*e.g.* $BaYCo_2O_{5.5}$ (DP) [3]), spin-state transition (*e.g.* $BaYCo_2O_{5.0}$ (DP) [4]) and simultaneous valence-separation and charge-ordering transition (*e.g.* $BaREFe_2O_{5.0}$ (DP) [5,6]).

The *A*-site ordered DP structure was first established for $BaRE(Cu_{0.5}Fe_{0.5})_2O_5$ with *RE* = Y in 1988 [7]. Later the same structure has been observed for compounds with various *RE*s and *B* constituents. The DP structure with the *B* site being occupied only by a single element is known for *B* = Co [8], Mn [9] and Fe [10]. For copper, the maximum *B*-site occupation so far reached is $x = 0.7$ in (Ba,La)Y$(Cu_xFe_{1-x})_2O_{5+\delta}$ by means of ultra-high-pressure high-temperature treatment [11]. Of $BaREB_2O_{5+\delta}$ (*B* = Co, Fe and Mn), the phases with *B* = Co are easiest to synthesize: they form in air, $O_2$ and $N_2$ atmospheres with most of the *RE*s and exhibit wide and tunable ranges of excess oxygen in the $REO_\delta$ layer [12,13]. The *B* = Fe phases require strongly reduced oxygen pressures to form, but allow continuous tuning of the oxygen nonstoichiometry [5,10]. Among the three systems, the last is the most challenging: successful synthesis of the *B* = Mn phases requires careful control of the redox conditions to avoid formation of stable binary oxides such as $BaMnO_{3-\delta}$ [14]. Furthermore, the oxygenation/deoxygenation behavior of the Ba*RE*$Mn_2O_{5+\delta}$ phases is not completely understood yet.

In the present contribution we present a three-step synthesis method to obtain a new intermediate-oxygen-content phase of $BaYMn_2O_{5.5}$ or $Ba_2Y_2Mn_4O_{11}$ consisting of (*i*) oxygen-getter-controlled low-$O_2$-pressure encapsulation synthesis of high-quality samples of $BaYMn_2O_{5+\delta}$ with $\delta \approx 0.0$, (*ii*) low-temperature post-annealing in $O_2$ to obtain the fully-oxygenated $\delta \approx 1.0$ compound, and (*iii*) subsequent annealing in $N_2$ at temperatures higher than 600 °C to finally reach the half-oxygenated $\delta \approx 0.5$ compound. Moreover, it is revealed that the excess-oxygen atoms are ordered in $BaYMn_2O_{5.5}$. (So far the half-oxygenated, oxygen-ordered Ba*RE*$Mn_2O_{5.5}$ phase has been reported only for the largest *RE* (= La) [15].)

## 2. Experimental

For the synthesis of $BaYMn_2O_{5+\delta}$ samples with the *A*-site ordered double-perovskite structure an oxygen-getter-controlled low-$O_2$-pressure encapsulation technique was first developed. As a precursor for the encapsulation synthesis we used a cation-stoichiometric mixture of $Y_2O_3$, $Mn_3O_4$ and $BaCO_3$ preheated in flowing $N_2$ gas at 1000 °C for 24 hours. The calcined powder mixture was pressed into pellets which were then placed in an evacuated silica ampoule together with FeO powder, the function of which was to act as a getter for excess oxygen. The synthesis was carried out at 1100 °C for 48 hours. Then the ampoule was quenched to room temperature. To obtain fully oxygenated $BaYMn_2O_{5+\delta}$, as-synthesized sample pellets were annealed in flowing $O_2$ gas at 600 °C for 48 hours. Intermediate-oxygen-stoichiometry samples were obtained by annealing the fully-oxygenated pellets in flowing $N_2$ gas at 800 °C for 48 hours. Both annealing treatments were carried out in a thermobalance (Perkin Elmer: Pyris 1 TGA) to *in situ* follow the changes occurring in the sample weight/oxygen content during annealing. The heating rate was 1 °C/min and the mass of the sample was *ca.* 100 mg. The oxygenation period at 600 $^{o}C$ was followed by slow cooling (with a rate of 1 $^{o}C/min$) down to room temperature, whereas after the deoxygenation treatment at 800 $^{o}C$ fast cooling (20 $^{o}C/min$) was employed. In order to better observe the oxygen incorporation/depletion characteristics parallel oxygenation/deoxygenation TG experiments were carried out for powderized samples of *ca*. 100 mg. In these experiments the isothermal heating period (of 48 hours) applied for the pellet samples at the final temperature was omitted.

The precise oxygen content was determined by iodometric titration for the three samples: as-synthesized, $O_2$-annealed and $N_2$-annealed $BaYMn_2O_{5+\delta}$. Parallel titration experiments were performed to warrant the obtained results. For the titration sample powder of *ca*. 20 mg was dissolved in deoxygenated 3 M HCl solution containing an excess of KI. The amount of $I_2$ liberated upon the reduction of higher-valent Mn species, *i.e.* $Mn^{III}$ and/or $Mn^{IV}$, to $Mn^{2+}$ ions was determined using 0.1 M $Na_2S_2O_3$ solution as the titrant. The titration experiment was carried out in $N_2$ atmosphere.

Magnetization of the samples was measured in a temperature range of 5 - 300 K under a magnetic field of 100 Oe using a SQUID magnetometer (Quantum Design: MPMS-XL5).

The lattice parameters and atomic positions of the samples were determined through Rietveld refinement (program Gsas) of synchrotron x-ray diffraction data collected at room temperature at SPring-8, Japan (beamline: BL02B2) using radiation with wavelength of 0.500981 Å. For these experiments the sample was crushed into fine powder and sealed in a quartz capillary of 0.1 mm in diameter. Taking an analogy to the case of oxygen-ordered $BaLaMn_2O_{5.5}$ [15], the present half-oxygenated $BaYMn_2O_{5.5}$ sample was indexed for an orthorhombic unit cell (space group *Ammm*) with the following relations to the simple perovskite unit cell; $a = a_p$, $b = 2b_p$ and $c = 4c_p$, where the values of the parameters are, $a$ = 3.771(1) Å, $b$ = 8.1590(1) Å and $c$ = 15.2709(3) Å. Very few faint reflections remained unindexed, the strongest among them being at $d$ = 3.060 Å, which could be attributed to a trace of impurity, $Y_2O_3$. Refinement of the data confirmed that the amount of $Y_2O_3$ in the sample was less than 3 wt-%. Preliminary electron diffraction (ED) data for the same half-oxygenated sample confirmed the superlattice due to oxygen ordering but showed even an additional set of weaker superspots for a repetition, $c = 8c_p$ [16]. Le Bail fitting of the x-ray diffraction pattern for such a supercell (assuming *Pmmm*) did not give any strong arguments for increasing the complexity of the structure model. The background was fitted with a 5th order polynomial (function #6 in Gsas). The peak profiles were described by pseudo-Voigt functions with additional corrections for peak asymmetry. Isotropic temperature factors were refined individually for the cations, whereas a single common factor was used for all the oxygen atoms. Table 1 provides the further details of the refinements.

## 3. Results and Discussion

The synchrotron x-ray diffraction data confirmed that the present oxygen-getter-controlled low-$O_2$-pressure encapsulation synthesis had yielded high-quality double-perovskite samples of $BaYMn_2O_{5+\delta}$. No signs of the commonly observed impurity phases, BaO, $Mn_2O_3$, MnO, $BaMnO_{3-\delta}$ and $YMnO_{3-\delta}$ [17,18], were detected. We expect that the oxygen partial pressure at the synthesis temperature, *i.e.* 1100 $^oC$, equilibrates at *ca.* 1.1 x $10^{-10}$ atm due to the redox couple of $FeO/Fe_3O_4$ [19]. For the as-synthesized sample, iodometric titration revealed an essentially stoichiometric, oxygen-deficient oxygen content, $\delta = -0.02(1)$.

Once the *A*-site ordered, oxygen-deficient (here $\delta \approx 0$) double-perovskite framework of $BaYMn_2O_{5+\delta}$ is formed under reduced oxygen partial pressures (here *ca.* 1.1 x $10^{-10}$ atm), the phase can be oxygenated by means of post-annealing in 1-atm $O_2$ atmosphere. In Fig. 1(a) the thermogravimetric (TG) curve for the oxygenation of as-synthesized $BaYMn_2O_{4.98}$ (powder) at 600 $^oC$ is shown. Upon heating the sample (at a rate of 1 $^oC$/min) incorporation of oxygen occurs in a single step in the temperature range of 250 - 400 $^oC$. The weight gain corresponds to ~0.9(1) oxygen atoms *per* formula unit, *i.e.* $\Delta\delta \approx 0.9(1)$. Iodometric titration revealed a value of 0.97(1) for the amount of excess oxygen, $\delta$, in the $O_2$-annealed sample. We may thus conclude in agreement with previous reports [17,20] that an essentially fully-oxygenated $BaYMn_2O_{5+\delta}$ (with $\delta \approx 1$) phase has been obtained from the oxygen-depleted $BaYMn_2O_{5+\delta}$ (with $\delta \approx 0$) sample by means of the low-temperature $O_2$ annealing.

In order to deoxygenate the fully-oxygenated $BaYMn_2O_{5.97}$ phase, the sample was annealed in a thermobalance in $N_2$ at 800 °C. From the observed weight loss the magnitude of $\Delta\delta$ was calculated at 0.5(1). In agreement with the TG result, iodometric titration revealed a value of $\delta$, 0.54(1), for the $N_2$-annealed sample. The TG curve obtained for the $N_2$ annealing of the powder sample is displayed in Fig. 1(b). It is seen that the weight loss occurs in a single sharp step around 550 - 650 °C. Attempts to obtain samples with $0.5 < \delta < 1$ by means of $N_2$ annealing at temperatures between 550 and 650 °C all ended up with depletion of oxygen down to the $\delta \approx 0.5$ level. This is in drastic contrast to the case of the $BaRECo_2O_{5+\delta}$ double perovskites [13] for which continuous tuning of oxygen nonstoichiometry within $0 < \delta < 0.7$ is possible by selecting a proper temperature for the $N_2$ annealing in the temperature range of 300 - 1000 °C, *i.e.* "temperature-controlled oxygen- depletion (TCOD)" [21] annealing. The result obtained for the present Mn-based double-perovskite samples indicates that the phases of $BaYMn_2O_{5+\delta}$ with the amount of excess oxygen, $\delta$, between 0.5 and 1.0 are not (easily) stabilized. Thus, only three distinct phases have been shown to exist in the $BaYMn_2O_{5+\delta}$ system: $BaYMn_2O_{5.0}$ ($BaYMn_2O_{4.98}$), $BaYMn_2O_{5.5}$ or $Ba_2Y_2Mn_4O_{11}$ ($BaYMn_2O_{5.54}$), and $BaYMn_2O_{6.0}$ ($BaYMn_2O_{5.97}$). Among these, the first [9,20] and the last [17,22] have been reported earlier whereas the one with the intermediate oxygen content is a new phase. In Fig. 2, we represent schematic illustrations of crystal structures of the three phases.

Synchrotron x-ray diffraction patterns for the samples are shown in Fig. 3. Refinement of the diffraction data confirmed the $P4/nmm$ [20] description for $BaYMn_2O_{5.0}$ with the lattice parameters, $a$ = 5.5494(1) Å and $c$ = 7.6523(1) Å, which are consistent with those previously reported [20]. Since atomic positions for the $BaYMn_2O_{6.0}$ phase had not been published, the unit cell was refined by Le Bail fitting based on the proposed monoclinic structure of space group $P2$ [22]. The derived values, $a = 5.5250(2)$ Å, $b = 5.5197(2)$ Å, $c = 7.6092(1)$ Å and $\beta = 90.294(2)^{o}$, are close to those reported in Ref. [22]. For the new $BaYMn_2O_{5.5}$ phase the diffraction pattern was compatible with space group *Ammm*, which had previously been revealed for $BaLaMn_2O_{5.5}$ [15]. The fitted pattern is shown in Fig. 4. Final atomic coordinates and displacement parameters are presented in Table 2 and the cation-oxygen bond distances and the corresponding bond-valence-sum values in Table 3. In Fig. 5, plotted are the lattice parameters and unit volume (as expressed in terms of the simple perovskite unit cell for convenience) for all the three phases against δ. As the amount of excess oxygen increases the unit volume gradually decreases. For $BaYMn_2O_{5.5}$ the lattice is clearly orthorhombic. For the Co-based analog, $BaYCo_2O_{5.5}$, both tetragonal and orthorhombic variants are reported [3], obtained by means of fast cooling and slow cooling, respectively. There the orthorhombic splitting is assumed to originate from ordering of excess oxygen atoms within the $YO_{0.5}$ layer [3]. Such ordering has recently been confirmed for orthorhombic $BaLaMn_2O_{5.5}$ [15].

For the $BaYMn_2O_{5.5}$ sample the Rietveld refinements gave an excellent fit to an orthorhombic structure with oxygen ordering (Fig. 4). In the crystal structure, barium and yttrium takes twelve and ten coordinated sites, respectively (Table 3). There are two types of isovalent manganese atoms; bond-valence-sum calculation revealed a valence value of 3.32/3.30 for Mn1/Mn2 (Table 3) in accordance with the expected oxidation state of III. The octahedral Mn2 site is strongly distorted, with roughly a (4 + 2) Mn-O coordination. This is in agreement with a $d^4$ Jahn-Teller state of HS $Mn^{III}$. The same applies for the square-pyramidal Mn1 site. Similar findings have been reported for $BaLaMn_2O_{5.5}$ [15].

The preliminary ED data for $BaYMn_2O_{5.5}$ [16] differ from those published for $BaLaMn_2O_{5.5}$ [15] by showing additional spots that correspond to a repetition $c = 8c_p$, *i.e.* twice the $c$ axis now adopted. In this respect we noticed that the refined common displacement factor for the oxygen atoms is unusually large. Refinements of individual displacement factors showed that O3 is behaving anomalously. One possibility is vacancies, which is, however, not supported by the result of iodometric titration analysis, giving rather a slightly oxygen-rich composition of $BaYMn_2O_{5.54}$. Although refinements of the O3 occupation number improved the *R*-factors significantly, this was considered to be an artifact since the thereby deduced oxygen content ($\delta$ = 0.32) was too low. Instead, an anisotropic description of the displacement parameters gave large values for U11 and U22. This indicates that the O3 atom may be described as a split atom in the *xy*-plane (oxygen content fixed to the value determined by iodometry); refined coordinates are (0.581, 0.053, 0.087) for a 25 % occupied 16-fold site. The present data do not allow further elaboration, but it is likely that the superstructure spots seen in the ED patterns actually reflect an ordered distribution of the *xy*-displacements now refined for the O3 atom. Structurally, the O3 distribution is of major importance for the Mn1 square-pyramidal coordination.

The results of magnetization measurements showed new features for $BaYMn_2O_{5.5}$ as compared with the earlier findings for both $BaYMn_2O_5$ and $BaYMn_2O_6$ (Fig. 6). The oxygen-free $BaYMn_2O_5$ has been reported to be a ferrimagnetic insulator below 165 - 167 K [9,20] exhibiting $Mn^{2+}/Mn^{3+}$ charge, orbital and spin orderings. For the present $BaYMn_2O_{4.98}$ sample, a ferrimagnetic transition is seen at 166 K; the observed magnetization of ~0.66 $\mu_B$ *per* Mn site is slightly larger than the previously observed value of 0.475 $\mu_B$ [20] and the expected value of 0.5 $\mu_B$ *per* formula Mn site. The fully-oxygenated $BaYMn_2O_6$ phase shows CE-type $Mn^{3+}/Mn^{4+}$ charge and orbital orderings at room temperature, and antiferromagnetic ordering below ~200 K according to a recent report [17]. For the present $BaYMn_2O_{5.97}$ sample an antiferromagnetic transition is determined at 179 K. Additionally an upturn in the magnetization curve is seen about 45 K. A similar but weaker feature is also visible in the data of Ref. [17], though not addressed by the authors. The magnetization curve for the $BaYMn_2O_{5.54}$ sample indicates a complex behaviour with two magnetic transitions about 133 K and 42 K. The higher-temperature transition is of ferro/ferrimagnetic type, corresponding to magnetization of 1.13 x$10^{-3}$ $\mu_B$ *per* Mn site. For the La-based oxygen-ordered compound, $BaLaMn_2O_{5.5}$, Caignaert *et al.* [15] concluded from neutron powder diffraction data an original magnetic structure with ferromagnetic spinladders along the

*b*-axis that are antiferromagnetically coupled along the *a*- and *c*-axes. The magnetic transition temperature was determined about 185 K. The lower-temperature transition for the $BaYMn_2O_{5.54}$ sample is akin to that seen for the $BaYMn_2O_{5.97}$ sample, suggesting that it may be originated from an impurity phase, such as $BaMnO_3$ [24] or $Ba_4Mn_3O_{10}$ [25], though not detected in diffraction patterns. We have initiated a neutron powder diffraction study to characterize the magnetic structures of the three $BaYMn_2O_{5+\delta}$ phases in detail and especially that of the new $\delta \approx 0.5$ phase.

## 4. Conclusion

We developed a three-step synthesis route to obtain a new oxygen and *A*-site cation ordered perovskite phase, $BaYMn_2O_{5.5}$. The route passed through intermediate steps where high-quality samples of the known $BaYMn_2O_{5.0}$ and $BaYMn_2O_{6.0}$ phases were yielded. For the new phase synchrotron x-ray diffraction analysis revealed an orthorhombic crystal structure and a long-range ordering of the excess oxygen atoms in the $YO_{0.5}$ layer giving rise to octahedrally and square-pyramidally coordinated trivalent manganese. In the magnetization experiments the phase showed a ferro/ferrimagnetic transition around 133 K.

## Acknowledgments

This work was supported by Grants-in-aid for Scientific Research (Nos. 15206002 and 15206071) from Japan Society for the Promotion of Science. Prof. Y. Moritomo and Dr. K. Kato are thanked for the synchrotron-radiation x-ray powder diffraction experiments which were performed at the Spring-8 BL02B2 beamline with approval of the Japan Synchrotron Radiation Research Institute (JASRI).

**Table 1.** Experimental conditions and relevant data for the Rietveld refinements of synchrotron powder x-ray diffraction data of $Ba_2Y_2Mn_4O_{11}$ ($BaYMn_2O_{5.5}$) at 298 K.

| **Parameter** | |
|---|---|
| Pattern range 2θ (°) | 3 - 40 |
| Step size Δ2θ (°) | 0.01 |
| Wavelength (Å) | 0.50098 |
| | |
| Space group | *Ammm* |
| *a* (Å) | 3.771(1) |
| *b* (Å) | 8.1590(1) |
| *c* (Å) | 15.2709(3) |
| *Z* ($Ba_2Y_2Mn_4O_{11}$) | 2 |
| V (Å$^3$) | 469.95(1) |
| No. observations | 3700 |
| No. reflections | 414 |
| No. refined params. | 21 |
| $R_{exp}$ | 0.029 |
| $R_{wp}$ | 0.064 |
| $R_F^2$ | 0.059 |

**Table 2.** Atomic coordinates for $Ba_2Y_2Mn_4O_{11}$ ($BaYMn_2O_{5.5}$) at 298 K as derived from Rietveld refinements of synchrotron powder x-ray diffraction data, calculated standard deviations in parentheses. Space group *Ammm*.

| **Atom** | **x** | **y** | **z** | **$U_{iso}$ (Å$^2$)** |
|---|---|---|---|---|
| Y | 0.5 | 0.2731(3) | 0 | 0.51(7) |
| Ba | 0.5 | 0.25 | 0.25 | 0.91(5) |
| Mn1 | 0.0 | 0.0 | 0.1145(3) | 0.10(5) |
| Mn2 | 0.0 | 0.0 | 0.3784(4) | 0.10(5) |
| O1 | 0.0 | 0.2228(12) | 0.1009(5) | 3.05(15) |
| O2 | 0.0 | 0.0 | 0.2525(17) | 3.05(15) |
| O3* | 0.5 | 0.0 | 0.0877(10) | 3.05(15) |
| O4 | 0.5 | 0.0 | 0.3843(13) | 3.05(15) |
| O5 | 0.0 | 0.0 | 0.5 | 3.05(15) |

* refined as split atom, 0.581(7), 0.053(2), 0.0874(9); occupation ¼. $U_{iso}$(oxygen) = 2.03(15) Å$^2$.

**Table 3.** Selected interatomic distances for $Ba_2Y_2Mn_4O_{11}$ ($BaYMn_2O_{5.5}$), calculated on the basis of results from Rietveld refinements of synchrotron powder x-ray diffraction data, calculated standard deviations in parentheses. The values for bond-valence-sums are calculated using the parameters of Ref. [23].

| Bond | Distance (in Å) | Bond Valence Sum |
|---|---|---|
| Y-O1 | 2.469(5) x 4 | |
| Y-O3 | 2.599(8) x 2 | 2.40 |
| Y-O4 | 2.559(14) x 2 | |
| Y-O5 | 2.643(1) x 2 | |
| | | |
| Ba-O1 | 2.965(6) x 4 | |
| Ba-O2 | 2.7783(4) x 4 | 2.47 |
| Ba-O4 | 2.893(15) x 2 | |
| | | |
| Mn1-O1 | 1.830(10) x 2 | |
| Mn1-O2 | 2.104 (26) | 3.32 |
| Mn1-O3 | 1.931(4) x 2 | |
| | | |
| Mn2-O1 | 2.285(10) x 2 | |
| Mn2-O2 | 1.915(26) | 3.30 |
| Mn2-O4 | 1.888(1) x 2 | |
| Mn2-O5 | 1.864(6) | |

## FIGURE CAPTIONS

**Fig. 1.** TG curves for (a) oxygenation of as-synthesized $BaYMn_2O_{4.98}$ in $O_2$, and (b) deoxygenation of oxygenated $BaYMn_2O_{5.97}$ in $N_2$. The precise oxygen content given for the starting material and the ending product of the TG annealing is a value obtained from iodometric titration. The $\Delta\delta$ value is calculated from the observed weight loss.

**Fig. 2.** Schematic illustration of the crystal structures of the three compounds, $BaYMn_2O_{5.0}$, $BaYMn_2O_{5.5}$ and $BaYMn_2O_{6.0}$.

**Fig. 3.** Synchrotron x-ray diffraction patterns for the three samples: $BaYMn_2O_{4.98}$, $BaYMn_2O_{5.54}$ and $BaYMn_2O_{5.97}$.

**Fig. 4.** Fitting of the synchrotron x-ray diffraction pattern for the $BaYMn_2O_{5.54}$ sample: observed (crosses) and calculated (solid line) intensities. The residuals between the observed and calculated intensities are shown in the bottom.

**Fig. 5.** Lattice parameters, $a_p$, $b_p$ and $c_p$, and the unit volume ($V_p$) of the simple perovskite unit cell plotted against oxygen content for the three $BaYMn_2O_{5+\delta}$ samples, $\delta$ = -0.02, 0.54 and 0.97. The following relation apply between the lattice parameters, $a$, $b$ and $c$, and the given simple-perovskite unit cell parameters: $a = b = \sqrt{2}a_p = \sqrt{2}b_p$ and $c = 2c_p$ for $BaYMn_2O_{4.98}$; $a = a_p$, $b = 2b_p$ and $c = 4c_p$ for $BaYMn_2O_{5.54}$; $a = \sqrt{2}a_p$, $b = \sqrt{2}b_p$ and $c = 2c_p$ for $BaYMn_2O_{5.97}$.

**Fig. 6.** Magnetization *versus* temperature behaviour for the three $BaYMn_2O_{5+\delta}$ samples, $\delta$ = -0.02, 0.54 and 0.97.

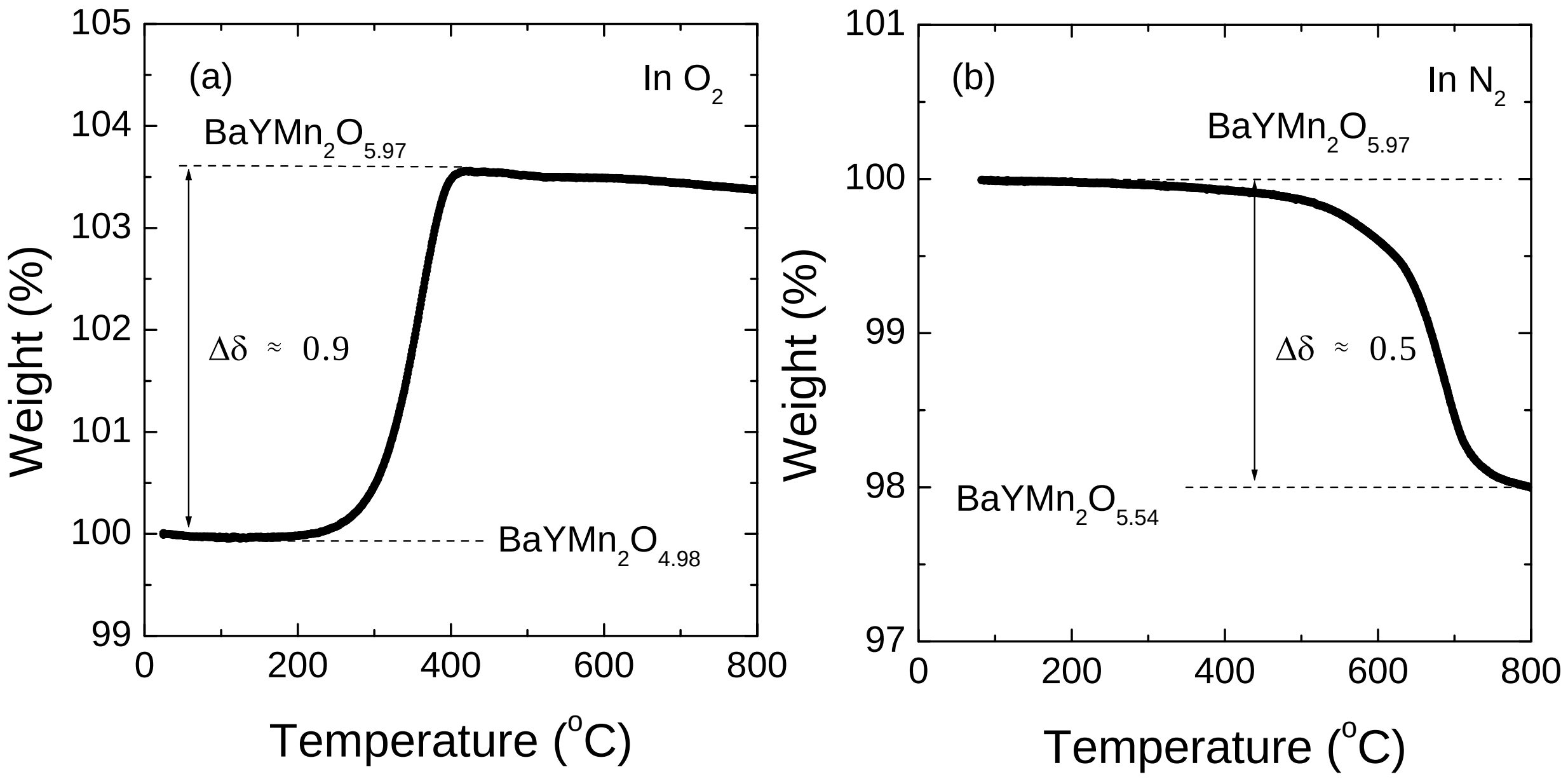

**Karppinen *et al:* Fig. 1.**

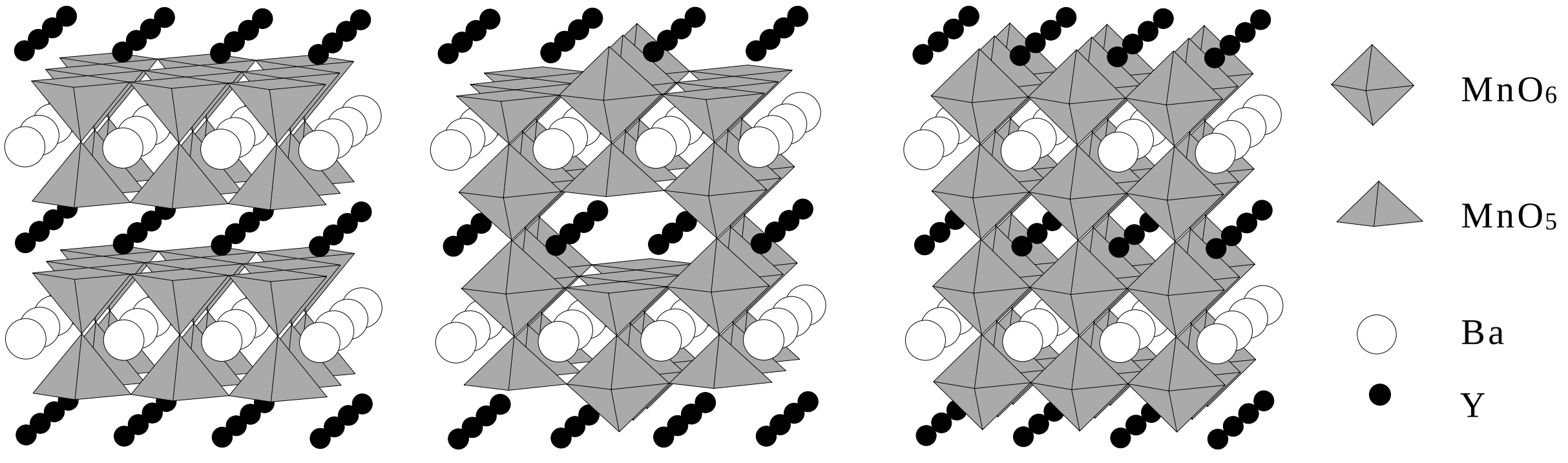

$BaYMn_2O_{5.0}$ $BaYMn_2O_{5.5}$ $BaYMn_2O_{6.0}$

**Karppinen *et al*: Fig. 2.**

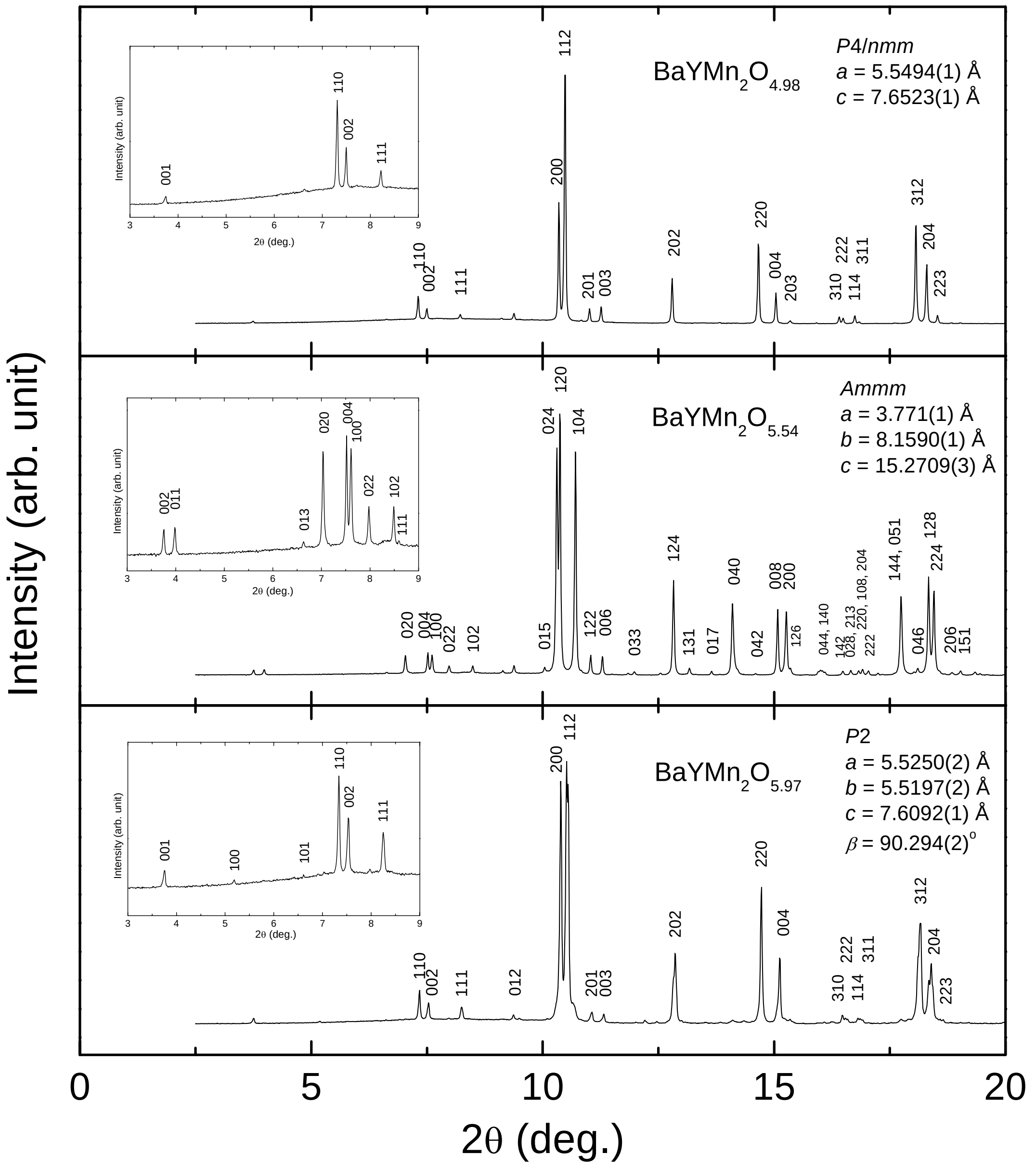

**Karppinen *et al*: Fig. 3.**

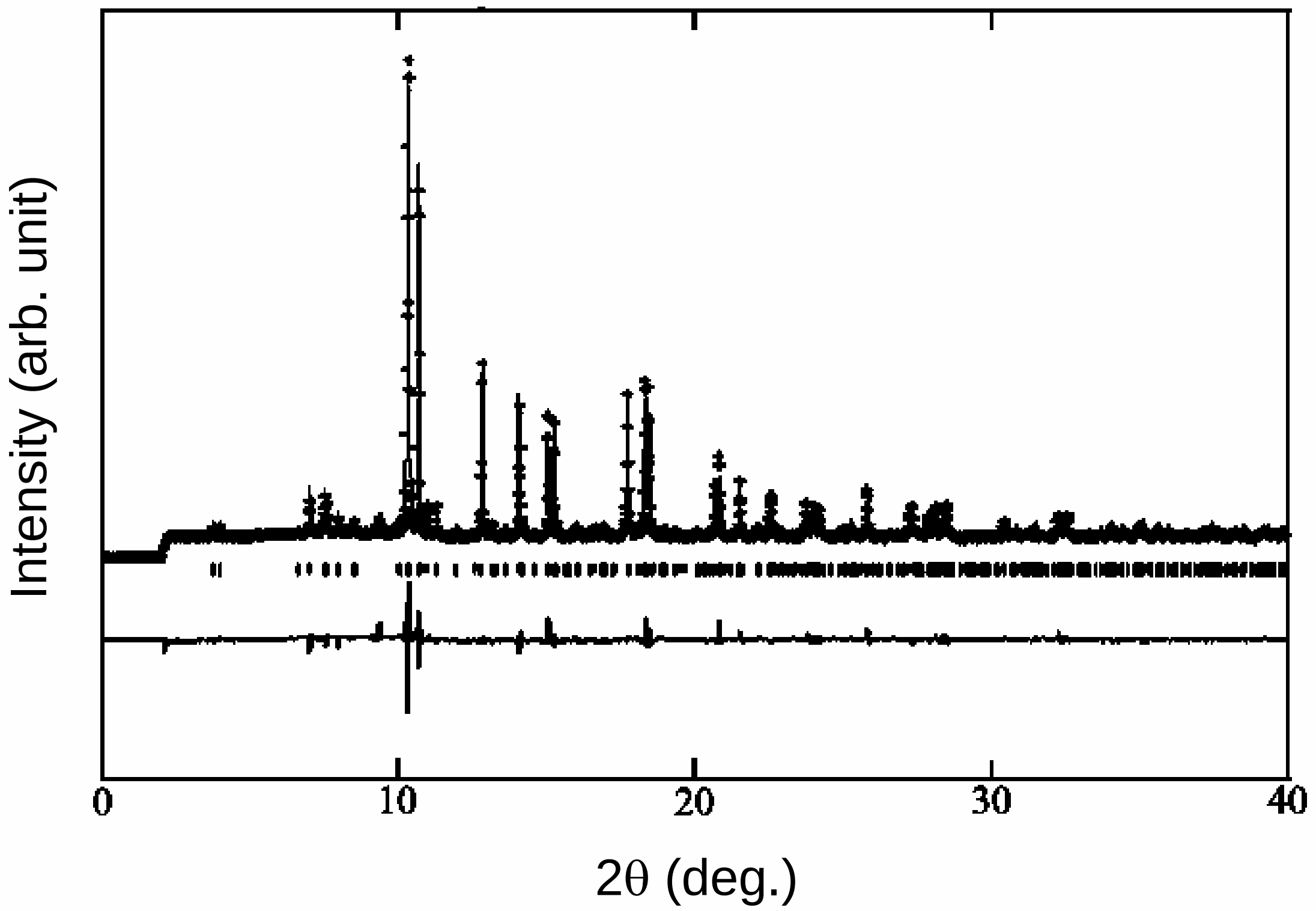

**Karppinen *et al*: Fig. 4.**

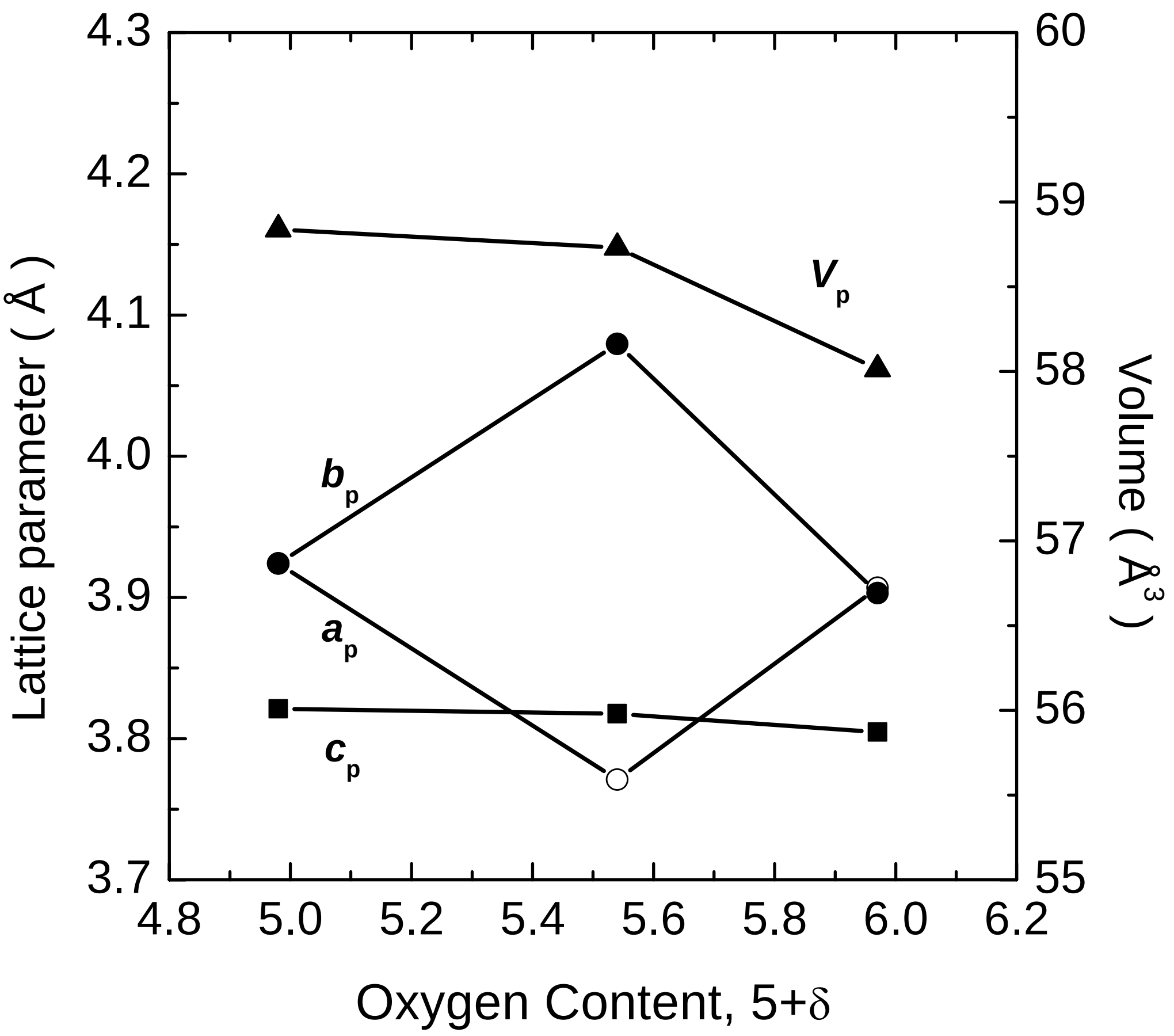

**Karppinen *et al*: Fig. 5.**

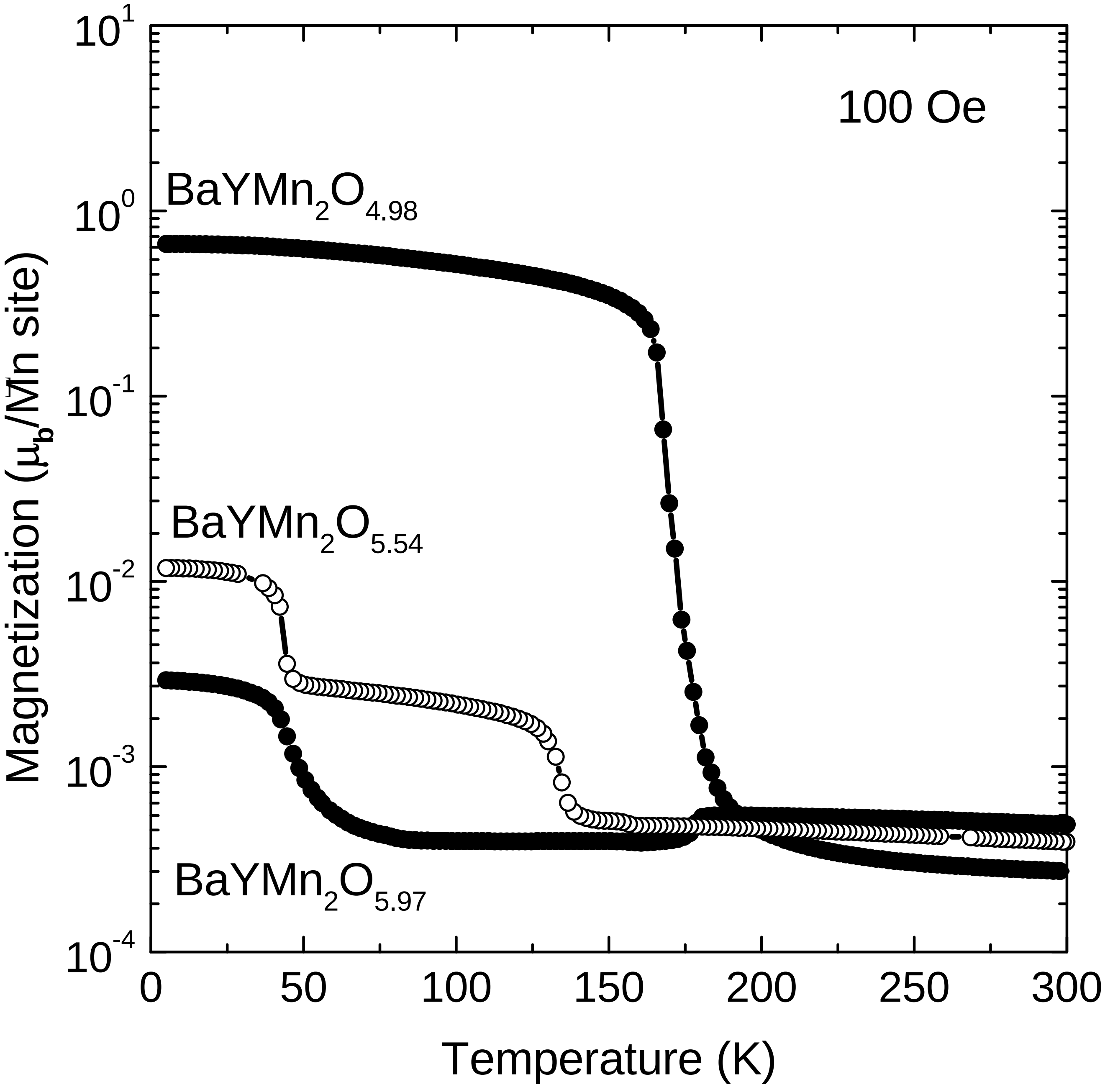

**Karppinen *et al*: Fig. 6.**